\documentclass[aps,preprint,preprintnumbers,nofootinbib,showpacs]{revtex4}
\usepackage{epsfig}
\usepackage{color}
\usepackage{amsmath}
\usepackage{amssymb}
\usepackage{dsfont}
\usepackage{mathtools}
\usepackage[normalem]{ulem}

\def\bea{\begin{eqnarray}}
\def\eea{\end{eqnarray}}
\def\sea{\nonumber \\&&}
\def\lla{\left\langle}
\def\rra{\right\rangle}
\def\za{\alpha}

\def\ssc{\scriptscriptstyle}
\def\lsim{\mathrel{\raise.3ex\hbox{$<$\kern-.75em\lower1ex\hbox{$\sim$}}} }
\def\gsim{\mathrel{\raise.3ex\hbox{$>$\kern-.75em\lower1ex\hbox{$\sim$}}} }
\def\pc{\psi_{c;jm}^n}
\def\ct{\cos{\theta}}
\def\sf{\sin{\phi}}
\def\cf{\cos{\phi}}
\def\st{\sin{\theta}}
\usepackage{graphicx,accents}
\usepackage{appendix}

\makeatletter
\DeclareRobustCommand{\cev}[1]{%
  \mathpalette\do@cev{#1}%
}
\newcommand{\do@cev}[2]{%
  \fix@cev{#1}{+}%
  \reflectbox{$\m@th#1\vec{\reflectbox{$\fix@cev{#1}{-}\m@th#1#2\fix@cev{#1}{+}$}}$}%
  \fix@cev{#1}{-}%
}
\newcommand{\fix@cev}[2]{%
  \ifx#1\displaystyle
    \mkern#23mu
  \else
    \ifx#1\textstyle
      \mkern#23mu
    \else
      \ifx#1\scriptstyle
        \mkern#22mu
      \else
        \mkern#22mu
      \fi
    \fi
  \fi
}

\makeatother

\begin{document}
\draft
\preprint{{\vbox{\hbox{NCU-HEP-k083}
\hbox{Nov 2019}
\hbox{rev. Dec 2019}
\hbox{ed. Feb 2020}
}}}
\vspace*{1.5in}

\title{ Analysis on Complete Set of Fock States with Explicit Wavefunctions for the Covariant Harmonic Oscillator Problem
\vspace*{.3in} }

\author{Suzana Bedi\'c} 

\affiliation{ICRANet, P.le della Repubblica 10, 65100 Pescara, Italy}
\affiliation{ICRA and University of Rome ``Sapienza'', Physics Department, P.le
A. Moro 5, 00185 Rome, Italy}

\author{Otto C. W. Kong,}
\email{otto@phy.ncu.edu.tw}

\affiliation{Department of Physics and 
Center for High Energy and High Field Physics,\\
National Central University, Chung-Li 32054, Taiwan}

\begin{abstract}\vspace*{.5in}
The earlier treatments of Lorentz covariant harmonic oscillator 
have brought to light various difficulties, such as reconciling 
Lorentz symmetry with the full Fock space, and divergence
issues with their functional representations. We present 
here a full solution avoiding  those problems. The complete 
set of Fock states is obtained, together with the corresponding 
explicit wavefunction and their  inner product integrals free 
from any divergence problem and the Lorentz symmetry fully 
maintained without  additional constraints imposed. By a simple 
choice of the pseudo-unitary representation of the underlying 
symmetry group, motivated from the perspective of the 
Minkowski spacetime as a representation for the Lorentz group, 
we obtain the natural non-unitary Fock space picture commonly 
considered though not formulated and presented in the careful 
details given here. From a direct derivation of the appropriate 
basis state wavefunctions of the finite-dimensional irreducible 
representations of the Lorentz symmetry, the relation
between the latter and the Fock state wavefunctions is also 
explicitly shown. Moreover, the full picture including the
states with non-positive norm may give consistent physics 
picture as a version of Lorentz covariant quantum mechanics. 
Probability interpretation for the usual von Neumann 
measurements is not a problem as all wavefunctions 
restricted to a definite value for the `time' variable are 
just like those of the usual time independent quantum 
mechanics. A further understanding from a perspective 
of the dynamics from the symplectic geometry 
of the phase space is shortly discussed.

\end{abstract}
\maketitle

\section{Introduction}
The importance of the harmonic oscillator problem 
in quantum mechanics can hardly be overstated. 
It is then easy to appreciate that the problem as 
formulated with the classical Minkowski spacetime 
instead of the Newtonian one as the starting point 
received a lot of attention since physicists started 
to think about `relativistic quantum mechanics'
 \cite{cho}. Here, we are talking about the latter 
relaxed from the usual textbook usage of the term. 
In fact, one may think about the quantum theory as found
in the textbooks becoming kind of standard only 
because of the failure to obtain a nice covariant 
formulation physicists would like to have. The 
covariant harmonic oscillator problem cannot avoid 
such a setting though. The obvious theoretical principle 
one would want to impose is covariance under the 
Lorentz symmetry $SO(1,3)$.
Perhaps we should make it clear that it is not our 
intention to fully address the general issue of the 
`relativistic' generalization of the standard quantum 
harmonic oscillator problem here. Nor do we want
to discuss about different formulations of `relativistic 
quantum mechanics'. Such tasks are certainly much
beyond our scope here. For that matter, there have 
been various different approaches, including the ones
in favor of going outside the framework of Lorentz 
or Poincar\'e symmetry \cite{11,12,13,14,15},
which are also of interest. For our case, it suffices 
to say that the big practical success of high energy physics under the framework of quantum field theory
certainly suggests the Lorentz covariant problem we
focus on here deserves serious studies. And there 
certainly has been no lack of efforts in that direction
from the beginning. The topic has been revisited 
more recently in Ref.\cite{B}. The noncompact 
nature of the Lorentz group leads to some quite 
nontrivial issues, as illustrated therein. 
We present here a full analysis on the
natural non-unitary Fock space formulation which 
gives the complete set of Fock states with explicit 
wavefunction solutions meeting the best expectations one could have for the Lorentz covariance feature. 
And there is no divergence in any of the wavefunctions 
or integrals for their inner products. This can only be 
achieved by giving up on the complete unitarity. 
We explain why replacing it with a pseudo-unitarity, 
reflecting the Minkowski instead of the Euclidean 
nature of the classical spacetime, is not only 
reasonable but desirable (see more in Ref.\cite{082}).

The consideration of unitarity as a necessary
requirement for quantum mechanics is tied to the
Born probability interpretation. There is, however, 
a very sensible way to look at quantum mechanics 
without the latter \cite{081,074}. Besides, 
a simple bottom line here is that even in the 
setting of quantum mechanics with the Copenhagen 
interpretation, the Born probability picture 
should not be strictly required to be extended 
to a spacetime description. Maintaining the total 
probability of finding a particle somewhere in the
space, at a particular moment of its existence, 
to be unity is one thing, asking for the total 
probability of finding a particle somewhere in the
spacetime to be unity is quite another. A von 
Neumann measurement of an observable for 
a spacetime wavefunction without specifying, 
or at least restricting, the time does not seem 
to be anything we can do anyway.

To lay the background for comparison, we summarize
here the key features of the usual unitary formulations
below. Readers may consult Ref.\cite{B} for details.

A naive formulation of a version of the quantum harmonic 
oscillator problem on the otherwise classical Minkowski
spacetime can be seen as a solution to the eigenvalue
equation
\bea \label{naive2}
\frac{1}{2\hbar} \left( \hat{X}_\mu \hat{X}^\mu +\hat{P}_\mu \hat{P}^\mu \right) 
  \psi_{\lambda}(x^\mu)
= {\lambda} \psi_{\lambda}(x^\mu)  \;,
\eea
with, in a direct analog to a three dimensional harmonic
oscillator problem, the position, $\hat{X}_\mu$, and the 
momentum, $\hat{P}_\mu$, operators represented by
$x_\mu$ and $- i\hbar \frac{\partial}{\partial x^\mu}$,
respectively, satisfying the commutation relation 
$[\hat{X}_\mu, \hat{P}_\nu] = i\hbar \eta_{\mu\nu}$, 
where $\eta_{\mu\nu}=\mbox{diag}\{-1,1,1,1\}$ 
is the Minkowski metric. 

The operator on the left hand side of Eq.(\ref{naive2}) 
can be written in terms of the Lorentz covariant 
ladder operators
\bea
\hat{a}_\mu=\hat{X}_\mu+i\hat{P}_\mu\;,
\qquad  \hat{a}^\dag_\mu=\hat{X}_\mu-i\hat{P}_\mu\;;
\qquad\left[\hat{a}_\mu,\hat{a}^\dag_\nu \right]= 2\hbar\eta_{\mu \nu}\;,
\eea
while the eigenfunctions $\psi_{\lambda}(x^\mu)$ 
correspond to the eigenstates of a (shifted) number 
operator. The $n$th-level  states are obtained by 
applying $n$ raising operators on a ground state, 
denoted $\left|0\rra$, annihilated by all the lowering 
operators. There is a freedom in the choice of operators 
to be taken as raising/lowering, giving rise to  
different Fock spaces. For a ground-state wavefunction 
$\lla x^\mu | 0 \rra \sim e^{\mp \frac{x_\mu x^\mu}{2\hbar}}$, 
annihilated by $\hat{a}_\mu$ ($\hat{a}^\dag_\mu$) 
for upper (lower) sign, to avoid divergence, one has 
to constrain the $x^\mu$ vectors to the spacelike
(timelike) domain. There is still a very tricky 
normalization problem. In fact, because of the infinite 
range of the boost parameter, the squared-integral 
norm still diverges and has to be handled somehow 
tactically (\emph{e.g.} by redefining the norm so to 
factor out that infinite volumn element). It is not
clear at all there can be a mathematically consistent
definition of the norm that leaves an interesting 
enough set of Fock states normalizable. Moreover, 
an abstract algebraic analysis yields many states with 
negative norm since, for the spacelike Fock space, 
$\lla 0 | \hat{a}_{\!\ssc 0} \hat{a}^\dag_{\!\ssc 0} | 0 \rra = -1$.
Similarly, $\lla 0 |  \hat{a}^\dag_i \hat{a}_i | 0 \rra = -1$
for the timelike case. With $\hat{X}_{\mu}$ and $\hat{P}_{\mu}$
defined Hermitian, all Lorentz transformations are
unitary, meaning preserving an inner product of
positive definite norms. That is in direct conflict
with the notion that the four $\hat{a}^\dag_\mu$, 
hence four $\hat{a}^\dag_\mu \left| 0 \rra$ 
states (or  four $\hat{{a}}_\mu \left| 0 \rra$ states),
should transform as a Minkowski four-vector. It is 
then not surprising at all that Bars \cite{B}
concluded that unitarity and covariance together
leave only Lorentz invariant states as admissible,
which is however really saying Lorentz symmetry
is completely trivialized, hence essentially not there.
The paper does give some results and discussion 
about the nonunitary Fock states though, only far from
the explicit full results we will present below.

One different way to obtain a unitary positively-normed 
space of states is to take as ground state the one
annihilated by $\hat{a}^\dag_{\!\ssc 0}$ and $\hat{a}_i$ 
($i=1,2,3$). All Focks states obtained from it have positive 
norms and states at each level $n$ form an infinite 
dimensional irreducible unitary representation of 
the Lorentz group. However, the ground state is not 
Lorentz invariant \cite{B}. Lorentz symmetry should 
hence be considered spontaneously broken, in 
contrast to our objectives.

The structure of the paper is the following. Sec. \ref{sec2}
 is dedicated to the pseudo-unitary representation for 
the Lorentz covariant harmonic oscillator problem. 
In \ref{2a} it is motivated from a parallel with the Minkowski 
spacetime representation of the Lorentz symmetry. \ref{2b} 
contains the explicit operator formulation and Fock states
wavefunctions, whose transformation properties under 
Lorentz boosts are illustrated in \ref{2c}. In \ref{2d} we 
present the Lorentz invariant pseudo-unitary inner product 
on the Hilbert space spanned by Fock states, in an algebraic 
as well as integral form, and elaborate further on the Lorentz
structure of the Hilbert space. Sec. \ref{sec3} gives a 
direct derivation of the functional form of the basis
states of finite dimensional irreducible representations 
of the Lorentz symmetry in relation to the problem,
and their explicit connection to the results in  \ref{sec2}. 
We address issues related to interpretations of the results
in  Sec. \ref{sec4}, before concluding in  Sec. \ref{sec5}.

\section{A Pseudo-Unitary Representation}\label{sec2}
\subsection{Motivation} \label{2a}

The Minkowski spacetime is a pseudo-unitary
irreducible representation of the Lorentz symmetry.
Its associated invariant is an indefinite vector norm 
of signature $(1,3)$. Each transformation acts on a 
four-vector as a (real) $SU(1,3)$ matrix. It is this 
pseudo-unitary representation that reduces properly 
back to the reducible $1+3$ dimensional representation 
of the Newtonian space and time. Such non-unitarity 
we see as the defining signature of spacetime physics.
The full $SO(3)$ invariant Fock space of the harmonic
oscillator serves well as a solid picture of the single 
particle phase space under quantum mechanics, 
especially under the serious treatment of rigged Hilbert 
space formulation \cite{rhs} giving full justice to the 
Hermitian nature of the position and momentum 
operators. A similar treatment of the $SO(1,3)$ version 
could play an equivalent role in the proper formulation 
of Lorentz covariant quantum mechanics { \cite{082}}. 
One way or another, the essence of going from 
Newtonian physics to `relativistic' physics should be 
{like} a direct consequence of embedding the Newtonian 
space and time into the Minkowski spacetime. It is then 
very desirable to have the  $SO(3)$ invariant Fock space, 
for the `three dimensional' quantum harmonic oscillator 
problem, sit inside a full $SO(1,3)$ invariant Fock space 
in a manner directly analogous to how the Newtonian 
space sits inside the Minkowski spacetime.

We want to have a complete set of Fock states with
sensible norms as solutions to the problem, keeping the
Lorentz symmetry while maintaining that there are
four $n=1$ states transforming as a Minkowski
four-vector. As an irreducible representation of the
Lorentz group, the latter corresponds to a non-unitary
one. But it is the same non-unitarity of the Minkowski
spacetime as a representation space. That is the 
natural framework to see the problem as a Lorentz 
covariant version of the rotational covariant 
picture of the `three dimensional' theory. 
The indefinite Minkowski norm is what is preserved 
by the Lorentz transformations. We seek its natural 
extension in the form of pseudo-unitary norm for 
the Fock space, upon the restriction to the subspace 
of the four $n=1$ states.

 \subsection{Operator representations and Fock states with Hermite polynomials}\label{2b}
We start at the level of symmetry or algebraic structure 
at the abstract level. The symbols  $X_\mu$, $P_\mu$, $a_\mu$,
$\bar{a}_\mu$, \dots etc. are to be seen as abstract algebraic
quantities, for which we seek a representation as operators on
a Hilbert space. The relevant Lie algebra is that of 
$H_{\!\ssc R}(1,3)$, given as
\bea &&
[J_{\mu\nu}, J_{\rho\sigma}] 
= i \hbar \left( \eta_{\nu\sigma} J_{\mu\rho} 
  + \eta_{\mu\rho} J_{\nu\sigma} - \eta_{\mu\sigma} J_{\nu\rho} 
   -\eta_{\nu\rho} J_{\mu\sigma}\right) \;,
\sea
[J_{\mu\nu}, X_\rho] =  i \hbar \left( \eta_{\mu\rho} X_{\nu} 
   - \eta_{\nu\rho} X_{\mu} 
  \right) \;,
  \sea
[J_{\mu\nu}, P_\rho] =  i \hbar \left( \eta_{\mu\rho} P_{\nu} 
   - \eta_{\nu\rho} P_{\mu} 
  \right) \;,
\sea
[X_\mu, P_\nu] =i\hbar \eta_{\mu\nu} I\;, \label{alg}
\eea
for which we focus on representations of the Lorentz
symmetry with vanishing spins, \emph{i.e.} its six 
generators $J_{\mu\nu} (\equiv -J_{\nu\mu})$
can be taken as $J_{\mu\nu}= X_\mu P_\nu - X_\nu P_\mu $. 
A unitary representation, as a direct extension of
the $H_{\!\ssc R}(3)$ case with only $X_i$ and $P_i$ 
for standard quantum mechanics, is straightforward
 \cite{Z,J}. Yet, at least when applied to the harmonic
oscillator problem, the Fock state wavefunctions 
have undesirable behavior and divergence unless 
restricted to spacelike or timelike domains, under
which there may be other mathematical issues for
the full theory. Besides, the integral inner products
in either case contain a divergent volume factor
that has to be artificially dropped for them to make
sense. This has been well analyzed in Ref.\cite{Z},
with their undesirable Lorentz transformation 
properties also well addressed in Ref.\cite{B}, as
summarized above.
The pseudo-unitary representation is obtained as
\begin{align} 
    X_i&\rightarrow \hat{X}_i & P_i&\rightarrow \hat{P}_i 
\nonumber\\
    X_{\!\ssc 0} &\rightarrow i\hat{X}_{\!\ssc 4} & P_{\!\ssc 0}&\rightarrow i \hat{P}_{\!\ssc 4} \; , \label{rep}
    \end{align}
where, as operators on a space of functions of real  variables
$x^a$ ($a=1,2,3,4$),
\begin{align}
    \hat{X}_a&=x_a, &  \hat{P}_a&=-i \hbar \frac{\partial}{\partial x^a} \equiv -i \hbar {\partial_{\!a}}\;.
\end{align}
We have $[\hat{X}_a, \hat{P}_b] =i\hbar \delta_{ab}$, with $\delta_{ab}$ 
being the Kronecker delta symbol. Note that  $X_{\!\ssc 0}$ 
and $P_{\!\ssc 0}$, and hence $J_{{\ssc 0}\nu}$,
are represented by anti-Hermitian operators, therefore 
we have a non-unitary representation of the group 
$H_{\!\ssc R}(1,3)$ or its subgroup $SO(1,3)$. 
For the complex combinations $a_\mu$ and 
$\bar{a}_\mu$, we have then
\begin{align}
\hat{a}_{\!\ssc 0}&=i\left(\hat{X}_{\!\ssc 4} +i \hat{P}_{\!\ssc 4} \right) 
=i\hat{a}_{\!\ssc 4}
\nonumber\\
 \hat{\bar{a}}_{\!\ssc 0}&=i\left(\hat{X}_{\!\ssc 4} -i \hat{P}_{\!\ssc 4} \right) 
=i\hat{a}_{\!\ssc 4}^\dag\;,\label{arep}
\end{align}
while $\hat{a}_a=\hat{X}_a+i \hat{P}_a$, $\hat{a}_a^\dag=\hat{X}_a-i \hat{P}_a$, 
satisfying $[\hat{a}_\mu,\hat{\bar{a}}_\nu]=2\hbar \eta_{\mu\nu}$ 
and $[\hat{a}_a,\hat{a}_b^\dag]=2\hbar \delta_{ab}$. 
The (total) number operator can be written as
\begin{equation}
    \hat{N}=\frac{1}{2\hbar}\eta^{\mu\nu}\hat{\bar{a}}_\mu \hat{a}_\nu
=\frac{1}{2\hbar}\delta^{ab}\hat{a}^\dag_a \hat{a}_{ b}^{} \;,
\end{equation}
and decomposed into a sum of the Hermitian number operators 
$\hat{N}=\hat{N}_{\!\ssc 0} + \hat{N}_{\!\ssc 1}  + \hat{N}_{\!\ssc 2}  + \hat{N}_{\!\ssc 3}$, 
where $\hat{N}_i=\frac{1}{2\hbar}\hat{a}_i^\dag \hat{a}_i^{}$ and
$\hat{N}_{\!\ssc 0}=-\frac{1}{2\hbar}\hat{\bar{a}}_{\!\ssc 0} \hat{a}_{\!\ssc 0}
=\frac{1}{2\hbar}\hat{a}_{\!\ssc 4}^\dag \hat{a}_{\!\ssc 4}=\hat{N}_{\!\ssc 4}$, 
easily seen from (\ref{arep}). We have
$[\hat{N}_{\!a}, \hat{a}_a] =-\hat{a}_a$ and 
$[\hat{N}_{\!a}, \hat{a}_a^\dag] =\hat{a}_a^\dag$.
The normalized Fock states are eigenstates of $\hat{N}_{\! a}$ operators,
\bea \label{Naeq}
 \hat{N}_{\!a} \left| n_{\ssc\! 1},n_{\ssc\! 2},n_{\ssc\! 3};n_{\ssc\! 4}\rra 
= n_a \left| n_{\ssc\! 1},n_{\ssc\! 2},n_{\ssc\! 3};n_{\ssc\! 4}\rra \;.
\eea
Solving (\ref{Naeq}) in $x^a$ coordinates, in which
\begin{align}
    \hat{N}_{\!a}&=\frac{1}{2\hbar}\left(\hat{X}_a^2+\hat{P}_a^2 \right)-\frac{1}{2} \nonumber\\
    &=\frac{1}{2\hbar}\left(x_a^2-\hbar^2  {\partial^2_{\!a}} \right)-\frac{1}{2}\;, \label{Nop}
\end{align}
we obtain the eigenstate wavefunctions as
\begin{gather}
 \lla x^a| n_{\ssc\! 1},n_{\ssc\! 2},n_{\ssc\! 3};n_{\ssc\! 4}\rra =
 \frac{1}{\pi\hbar} e^{ -\frac{x_a x^a}{2\hbar}} 
  \Tilde{H}_{\!n_1}\!\!\left(\frac{x^{\ssc 1}}{\sqrt{\hbar}}\right)
  \Tilde{H}_{\!n_2}\!\!\left(\frac{x^{\ssc 2}}{\sqrt{\hbar}}\right) 
  \Tilde{H}_{\!n_3}\!\!\left(\frac{x^{\ssc 3}}{\sqrt{\hbar}}\right)
  \Tilde{H}_{\!n_4}\!\!\left(\frac{x^{\ssc 4}}{\sqrt{\hbar}}\right) \;,
\label{Herm}
\end{gather}
where $\Tilde{H}_{\!n_a}= [2^{n_a} n_a!]^{-\frac{1}{2}} H_{\!n_a}$, 
$H_{\!n_a}\!\!\left(\frac{x^{a}}{\sqrt{\hbar}}\right)$
being the standard Hermite polynomials. So, we have an explicit
solution for a complete set of the Fock states wavefunctions 
without any problem of the other formulations.  

In terms of $\hat{X}_a$ and  $\hat{P}_a$, the above
is just like a quantum version of  harmonic oscillator 
in the four Euclidean classical dimensions. The Hilbert 
space spanned by the eigenstate wavefunctions looks
completely conventional with an inner product 
giving a positive definite norm for the eigenstates
in a usual manner. However, we only have to introduce 
the notation $\hat{X}_{\!\ssc 0}=i\hat{X}_{\!\ssc 4}$ and
$\hat{P}_{\!\ssc 0}=i\hat{P}_{\!\ssc 4}$ to see
that $(\hat{N} +2) = \frac{1}{2 \hbar} \eta^{\mu\nu} 
 \left (\hat{X}_\mu \hat{X}_\nu + \hat{P}_\mu \hat{P}_\nu \right)$
corresponds exactly to the naively expected Hamiltonian
of the covariant harmonic oscillator in Eq.(\ref{naive2}).
It is interesting to note that identifying $\hat{X}_{\!\ssc 0}$
simply as $-\hat{X}_{\!\ssc 4}$ (and $\hat{X}^{\!\ssc 0}$
as $\hat{X}^{\!\ssc 4}$), and the same for $\hat{P}_{\!\ssc 0}$, 
works too though the Hermitian $\hat{X}_{\!\ssc 0}$ 
and $\hat{P}_{\!\ssc 0}$ then differs from the 
representations of ${X}_{\!\ssc 0}$ and ${P}_{\!\ssc 0}$ 
with an $i$ factor. The non-unitary nature of the 
representation and a sensible notion of a pseudo-unitary 
inner product on the Hilbert space can be seen by looking 
into the eigenstates and their transformation properties 
under the Lorentz symmetry, which we turn to next.

\subsection{Transformation properties under the Lorentz boosts}\label{2c}

The Lorentz-algebra generators $J_{\mu \nu}$ 
are represented by the operators 
$\hat{J}_{\mu \nu}=\hat{X}_\mu \hat{P}_\nu-\hat{X}_\nu \hat{P}_\mu$, 
where $\hat{J}_{\! ij}$ form a usual, unitarily represented  
$SO(3)$ subalgebra of spatial rotations, while 
\begin{align}
\hat{J}_{{\ssc 0}  i}&= \hat{X}_{\!\ssc 0} \hat{P}_{ \! i}-\hat{X}_{ \! i} \hat{P}_{\!\ssc 0}
=i \left( \hat{X}_{\ssc \! 4} \hat{P}_{\!i}-\hat{X}_{ \! i} \hat{P}_{\ssc \! 4} \right)\;
\end{align}
are the anti-Hermitian boost operators. To examine the nature 
of the obtained states under the Lorentz transformation, we act 
with the boost generator in the, arbitrarily chosen, $x^{\ssc 3} $ 
direction on the eigenstate  (\ref{Herm}). 
Using the properties of Hermite polynomials we get
\footnote{Note that the coefficients differ from the  erroneous ones in the published version of the paper \cite{83sym}.}
\begin{align} \label{infboost3}
    \hat{J}_{ \ssc 03} \left| n_{\ssc\! 1},n_{\ssc\! 2},n_{\ssc\! 3};n_{\ssc\! 4}\rra
  =\hbar\left(  \sqrt{n_{\ssc\! 3}(n_{\ssc\! 4}+1)} 
    \left| n_{\ssc\! 1},n_{\ssc\! 2},n_{\ssc\! 3}\!-\!1;n_{\ssc\! 4}\!+\!1\rra - \sqrt{n_{\ssc\! 4}(n_{\ssc\! 3}+1)} 
  \left| n_{\ssc\! 1},n_{\ssc\! 2},n_{\ssc\! 3}\!+\!1;n_{\ssc\! 4}\!-\!1\rra\right)\,.
\end{align}
We look into $n=1$ level, as those four states should correspond
to the components of a four vector. From (\ref{infboost3}) we 
can obtain $\hat{J}_{ \ssc 03}$ as a matrix
 \begin{equation}
     \hat{J}_{ \ssc 03} =\hbar \begin{pmatrix}
0&0&0&0\\
0&0&0&0\\
0&0&0&-1\\
0&0&1&0\\ 
\end{pmatrix}\;.
\end{equation}
Exponentiating, we get
\begin{equation}
    e^{i \frac{\za}{\hbar} \hat{J}_{ \ssc 03}}
= \begin{pmatrix}
1&0&0&0\\
0&1&0&0\\
0&0&\cosh{\za}& -i\sinh{\za}\\
0&0&i\sinh{\za}&\cosh{\za} \;
\end{pmatrix}\equiv \hat{\Lambda}_{\ssc 03}\,,
\end{equation}
the corresponding finite boost by the real parameter $\za$. 
Alternatively, we can see the transformation as a rotation 
in $x^{\ssc 3}\mbox{-}x^{\ssc 4}$ plane by a purely imaginary 
angle $i \za$. We find the action of $\hat{\Lambda}_{\ssc 03}$ 
on arbitrary function $f(x^{a})$ as
\begin{equation}\label{repg}
\left(\pi(\hat{\Lambda}_{\ssc 03})f\right)(x^{a}) 
= f \left(\hat{\Lambda}_{\ssc 03}^{-1}x\right)
= f \left(x^{\ssc 1},x^{\ssc 2},\cosh\za\,x^{\ssc 3}+i \sinh{\za}\,x^{\ssc 4},
   -i \sinh{\za}\,x^{\ssc 3}+\cosh\za\,x^{\ssc 4} \right)\,.
\end{equation}
In particular,
\begin{align}
    \pi(\hat{\Lambda}_{\ssc 03}) \lla x^a| 0,0,1;0\rra 
&= \frac{1}{\pi\hbar} e^{ -\frac{x_a x^a}{2\hbar}} 
 \Tilde{H}_{\!\ssc 1}\!\!\left(\frac{\cosh\za\,x^{\ssc 3}+i \sinh{\za}\,x^{\ssc 4}}{\sqrt{\hbar}}\right)   
\nonumber\\
 &=\cosh\za  \lla x^a| 0,0,1;0\rra+i \sinh{\za}   \lla x^a| 0,0,0;1\rra\;,
\end{align}
\begin{align}
    \pi(\hat{\Lambda}_{\ssc 03}) \lla x^a| 0,0,0;1\rra 
&= \frac{1}{\pi\hbar} e^{ -\frac{x_a x^a}{2\hbar}} 
 \Tilde{H}_{\!\ssc 1}\!\!\left(\frac{-i \sinh{\za} \,x^{\ssc 3}+ \cosh{\za}\,x^{\ssc 4}}{\sqrt{\hbar}}\right)   
\nonumber\\
 &=- i \sinh{\za} \lla x^a| 0,0,1;0\rra+\cosh{\za}   \lla x^a| 0,0,0;1\rra\;,
\end{align}
while $\lla x^a| 1,0,0;0\rra$ and $\lla x^a| 0,1,0;0\rra$
are invariant as $f(x_{a}x^{a})$ is obviously invariant 
under any Lorentz transformation.

Seen differently, we can introduce 
$|n\rangle_{_{\!\ssc 0}} \equiv  \left| n_{\ssc\! 0};n_{\ssc\! 1},n_{\ssc\! 2},n_{\ssc\! 3}\rra
= (i)^{n_{4}}\left| n_{\ssc\! 1},n_{\ssc\! 2},n_{\ssc\! 3};n_{\ssc\! 4}\rra$, 
with $n_{\ssc\! 0}=n_{\ssc\! 4}$, to show that in the basis 
formed by four $|n=1\rangle_{_{\!\ssc 0}}$ states, 
$\hat{\Lambda}_{\ssc 03}$ takes the usual form 
\begin{equation} \label{exmv}
 \begin{pmatrix}
\cosh\za&0&0& \sinh \za\\
0&1&0&0\\
0&0&1&0\\
 \sinh \za&0&0&\cosh \za
\end{pmatrix}\;,
\end{equation}
preserving a Minkowski norm on the real span of
the four $|n=1\rangle_{_{\!\ssc 0}}$ vectors. 
The states hence transform as components of 
a Minkowski four-vector. In fact, that real
span can actually be seen as a model of the
Minkowski spacetime with the $SO(3)$ invariant
subspace spanned by the single $n_{\ssc\! 4}=1$ state
and the complementary subspace spanned by the three 
$n_{\ssc\! 4}=0$ states, modeling the Newtonian
time and space, respectively.

The Minkowski norm, or the extension of it to
 the complex span of the $|n=1\rangle_{_{\!\ssc 0}}$ 
vectors, and further to the whole Hilbert space
spanned by all the Fock states, is definitely not
unitary. We seek exactly an inner product, or 
rather an invariant bilinear functional \cite{gf5}, 
different from the standard $\lla \phi | \phi' \rra$
corresponding to the $L^2$-norm for the 
wavefunctions, one that is invariant under 
any Lorentz transformation. 


\subsection{The pseudo-unitary inner product or invariant bilinear functional}\label{2d}
Fock states wavefunctions, given in Eq.(\ref{Herm}), 
are orthonormal according to 
$\int \lla  n_{\ssc\! 1},n_{\ssc\! 2},n_{\ssc\! 3};n_{\ssc\! 4}|x^a\rra 
   \lla x^a | m_{\ssc\! 1},m_{\ssc\! 2},m_{\ssc\! 3};m_{\ssc\! 4}\rra \,d^4x =\delta_{nm}$,
as the inner product is usually defined on a unitary 
Hilbert space. Label $n$ here is to be understood as 
$\left(n_{\ssc\! 1},n_{\ssc\! 2},n_{\ssc\! 3};n_{\ssc\! 4}\right)$, 
and similar for $m$. Therefore, we have  $\lla n|m \rra=\delta_{nm}$. 
Since the Lorentz transformations, boosts in particular, 
are not represented by unitary operators, such an 
inner product cannot be preserved in general, as can 
easily be seen from the results above, {\em e.g.} we have
\begin{equation}
  \lla  \hat{\Lambda}_{\ssc 03}\left(0,0,1;0\right)| \hat{\Lambda}_{\ssc 03}\left(0,0,1;0\right)\rra
=\cosh^2\za + \sinh^2\za \neq 1\;.
\end{equation}
Instead, we define another inner product given 
through the Fock state basis as
\begin{align}\label{2ndip}
     \lla \!\lla n| m \rra\!\rra &
=(-1)^{n_{\ssc 4}} \lla n_{\ssc\! 1},n_{\ssc\! 2},n_{\ssc\! 3};n_{\ssc\! 4}|m_{\ssc\! 1},m_{\ssc\! 2},m_{\ssc\! 3};m_{\ssc\! 4} \rra=(-1)^{n_{\ssc 4}} \delta_{nm}\;,
\end{align}
and extend it to the full vector space assuming sesqulinearity.
It gives an indefinite norm, which is the natural extension
of the Minkowski norm on the subspace of the real span
of the four $|n=1\rangle_{_{\!\ssc 0}}$ vectors, and is 
invariant under the Lorentz transformations. In particular, 
for the boost $\hat{\Lambda}_{\ssc 03}$ we have
\begin{equation}
         \lla\!\!\lla \hat{\Lambda}_{\ssc 03}(0,0,1;0)|\hat{\Lambda}_{\ssc 03}(0,0,1;0)\rra\! \!\rra 
=  \lla \!\lla0,0,1;0|0,0,1;0 \rra \!\rra=1\;,
\end{equation}
\begin{equation}
         \lla\!\!\lla \hat{\Lambda}_{\ssc 03}(0,0,0;1)|\hat{\Lambda}_{\ssc 03}(0,0,0;1)\rra\! \!\rra 
=  \lla \!\lla 0,0,0;1|0,0,0;1 \rra\! \rra=-1\;.
\end{equation}
Moreover, a state vector that is proportional to
the sum or difference of the two states here above
have zero norm under the inner product. We have
in general spacelike, timelike, and lightlike 
state vectors with positive, negative, and vanishing
pseudo-unitary norms, respectively. All vectors
have finite norm and are all normalizable to
the norm values of 1, 0, and -1, though the notion
of normalization is an empty one for the lightlike
states obviously. It is important to note that
normalizations with respect to $\lla\cdot|\cdot\rra$
and $\lla\!\lla\cdot|\cdot\rra\!\rra$ are 
in general not the same. All the basis Fock states
are however normalized with respect to both and
none of the basis states is lightlike.

Splitting the pseudo-unitary inner product notation
$\lla\!\lla \phi | \phi' \rra\!\rra$, one should consider
the ket $\left.\left| \phi' \rra\!\rra$ as simply
another notation for $\left| \phi' \rra$, while 
the bra $\lla\!\lla \phi \right|\right.$ as a
linear functional is in general different from
$\lla \phi \right|$. We have explicitly
$\lla\!\lla n \right|\right.= (-1)^{n_{\ssc 4}}  \lla n\right|$
which defines all $\lla\!\lla \phi \right|\right.$ 
implicitly. We have then for the inner product
\bea
\lla\!\lla \phi | \phi' \rra\!\rra
&& = \sum_n \lla\!\lla \phi |n \rra \!\lla n | \phi' \rra\!\rra
= \sum_n \lla\!\lla \phi |n \rra\!\rra  \lla n | \phi' \rra
\sea
=  \sum_n \overline{\lla\!\lla n|\phi  \rra\!\rra}  \lla n | \phi' \rra
= \int d^4x \sum_n  \lla \phi | n \rra (-1)^{n_{\ssc\! 4}}  \lla n | x^a \rra  \lla x^a  | \phi' \rra 
\sea
= \int d^4x \sum_n  \lla \phi | n \rra  \lla n | x^i,-x^{\ssc\! 4} \rra  \lla x^a  | \phi' \rra 
\sea
= \int d^4x  \lla \phi  | x^i,-x^{\ssc\! 4} \rra  \lla x^a  | \phi' \rra \;,
\eea
where we have used the fact that the wavefunctions
$\lla x^a | n \rra$, given explicitly in Eq.(\ref{Herm}), 
are odd and even in $x^{\ssc\! 4}$ for odd and even 
${n_{\ssc\! 4}}$, respectively. This gives a nice 
integral representation of it in terms of the  
wavefunctions\footnote{
In fact, in some sense, it may be more proper 
to write things in terms of an alternative 
formulation of the wavefunctions as 
$\lla\!\lla x^a|\phi  \rra\!\rra 
=  \sum_n (-1)^{n_{\ssc\! 4}}  \overline{\lla\!\lla n |x^a  \rra\!\rra} {\lla\!\lla n|\phi  \rra\!\rra}
= \sum_n (-1)^{n_{\ssc\! 4}} {\lla x^a  | n \rra} {\lla n|\phi  \rra}$.
The latter is however a lot more clumsy to work with.
Moreover, having two wavefunction representations
of the states here only causes potential confusion.
}.
Note that on the Hilbert space for a non-unitary 
representation of a noncompact group, there may not 
exist an invariant inner product. Certainly not 
a positive definite one.  The wavefunctions of the 
states may not be squared integrable either. The 
appropriate structure to look for is an invariant 
bilinear functional \cite{gf5}. Our 
$\lla\!\lla \phi | \phi' \rra\!\rra$ inner
product is exactly a gadget of that kind.

There is a simple way to write the mathematical
relation between the two inner products that gives
also an easy way to see the Lorentz invariance
of the pseudo-unitary one. It is given in terms
of a parity operator $\mathcal{P}_{\!\ssc 4}$, which
sends $x^{\!\ssc 4}$ to $-x^{\!\ssc 4}$, as
\bea \label{2ndipP4}
\lla\!\lla \phi | \phi' \rra\!\rra
= \lla \phi | \mathcal{P}_{\!\ssc 4} | \phi' \rra \;.
\eea
We can actually take this as the definition. The 
$(-1)^{n_{\!\ssc 4}}$ factor in our definition of the 
inner product  in term of the Fock state basis above
is exactly the $\mathcal{P}_{\!\ssc 4}$ eigenvalue 
of $\left|n\rra$. With it, we have nicely
\bea
\lla\!\lla \phi | \phi' \rra\!\rra
= \int d^4x  \lla \phi | \mathcal{P}_{\!\ssc 4}| x^a \rra   \lla x^a  | \phi' \rra 
= \int d^4x  \lla \phi  | x^i,-x^{\ssc\! 4} \rra  \lla x^a  | \phi' \rra \;.
\eea

A good way to appreciate the Lorentz
structure of the Hilbert space spanned
by the Fock states is the following. We
first look at the parallel for the case of the 
`three dimensional' quantum harmonic
oscillator. The three $n=1$ states transform
under $SO(3)$ as components of an 
Euclidean three-vector. The $n=0$ state
is invariant. The two constant $n$-level
subspaces are vector spaces for the three
dimensional defining representation and 
the trivial representations of $SO(3)$. For
the $n=2$ level, it corresponds exactly 
to the symmetric part of the product 
of two $n=1$ representations, {\em i.e.}
transforming as the Euclidean symmetric 
two tensor and the invariant ($n=0$). 
The standard $n=2$ wavefunctions 
clearly show that, for an explicit check.
One goes on to the higher $n$-tensors 
for the higher $n$ levels. As also similarly
discussed in Ref.\cite{B}, actually for the general
Minkowski case, at the $n$ level, the full set 
of Fock states is a symmetric tensor of $SU(1,3)$ 
which reduces to irreducible representations 
of $SO(1,3)$ corresponding to the rank of the 
traceless tensors in the decomposition. The rank
numbers are $n$, $n-2$, \dots, (0 or 1). The 
pattern is essentially the same for any `$l+m$ 
dimensional' harmonic oscillator with the Fock 
states at level $n$ obtained by the action of 
$n$ creation operators on the $n=0$ state, 
with the $l+m$ independent creation operators 
transforming as a $(l+m)$-vector.  The structure 
is not sensitive to the actual background 
signature $(l,m)$ the latter has. Such 
representations, of $SU(l,m)$ or $SO(l,m)$, 
are unitary only for the Euclidean case. For 
`three dimensional' case, the rank of each of
those traceless (Cartesian) tensors is exactly 
the $j$ value. Back to our $1+3$ Lorentzian
case, the finite dimensions of those traceless
irreducible tensors are given by the square of 
rank plus one, with the full result explicitly 
illustrated in the next section. The way the Fock 
states for the `three dimensional' states sit 
inside our Fock states at each $n$-level can 
also be easily traced from the perspective
of the Cartesian tensors. 

The nature of the $n$-level states as components
of the symmetric $n$-tensors can also be directly
seen by looking at the wavefunctions given
in terms of products of the Hermite polynomials 
with the common invariant factor $e^{-\frac{x_ax^a}{2\hbar}}$,
which is essentially the $n=0$ state wavefunction.
It is then easy to appreciate the right pseudo-unitary
inner product as to be given by Eq.(\ref{2ndip}) 
or  Eq.(\ref{2ndipP4}). The norm as an invariant
should better be expressed as 
$\int d^4x \lla \phi|x_a\rra\!\lla x^a | \phi \rra$
so that the upper indices in the wavefunction
$\lla x^a | \phi \rra$ can be contracted with the 
lower indices in the otherwise conjugate function 
$\lla \phi|x_a\rra$. For an Euclidean case $x_a=x^a$, 
as for the unitary inner product. To get to the 
pseudo-unitary inner product which goes 
along with the Minkowski nature of the tensors, it 
is then obvious that we only need to turn the 
$x_{\!\ssc 4}$ variable appearing in $\lla \phi|x_a\rra$,
which are the tensors with lower indices, into
$-x^{\!\ssc 4}$ or $\eta_{\!\ssc 00} x^{\!\ssc 4}$. 
The extra $i$ factor involved in the exact state 
for the $n=1$ level as the component of a 
Minkowski four-vector, as discussed right above 
and in relation to Eq.(\ref{exmv}), does not matter 
due to the sequlinearity of the inner product.
The invariant factor $e^{-\frac{x_ax^a}{2\hbar}}$
of course does not change, though it is to be
interpreted as $e^{-\frac{1}{2\hbar}\sum (x^a)^2}$ and 
$e^{-\frac{1}{2\hbar}(-x_{\!\ssc 4})^2 -\frac{1}{2\hbar}\sum (x_i)^2}$, 
accordingly.

\section{Fock states as representations of Lorentz symmetry}\label{sec3}
Our final task is to relate our Fock states to the basis 
states of the irreducible representations of the Lorentz 
symmetry explicitly derived. Since the number operator 
$\hat{N}$ commutes with Lorentz transformations, the 
collection of Fock states at a fixed level $n$ spans a 
Lorentz representation, generally reducible into a sum 
of finite-dimensional irreducible ones. Noncompactness 
of the Lorentz group implies the non-unitarity of the latter. 
They can be labeled \cite{Ba}  (see also Ref.\cite{gf5}) by 
two independent numbers $(j_o,c)$, the integer or 
half-integer $j_o$, corresponding to the spin, and the 
complex number $c$ characterizing the spin independent
Casimir invariant. Since our problem at hand is spinless, 
we simply drop the vanishing $j_o$. A convenient labeling 
of basis states is given by $\left|j_o,c;j,m\rra$, hence 
$\left|c;j,m\rra$, which transform as the familiar angular 
momentum states $\left|j,m\rra$ under the $SO(3)$ 
subgroup. For finite dimensional representations, which we 
are interested in, $c$ is a natural number and  $j=0,1,...,c-1$. 
$c=1$ is a trivial representation. All the others are
nonunitary. The smallest nontrivial one, $c=2$,
is then a sum of $j=0$ and $j=1$ representations
of $SO(3)$, which is the complex extension of
the one for the Minkowski spacetime. The 
dimensions of such irreducible representations 
are simply given by $c^2$.

To find the explicit formulation of Fock states in terms 
of the irreducible Lorentz representations we first solve 
the relevant differential equations for the latter, obtaining 
the corresponding functions in a coordinate form, and
show the way to obtain any Fock state as their linear 
combinations. The result is completely in accordance 
with discussions in the last part of the previous section.

The basis functions for irreducible Lorentz representation 
can be found as solutions of the eigenvalue equations of  
the nonzero Lorentz algebra Casimir operator $\hat{C}$, 
and the standard spherical harmonic in three dimensions. 
Additionally, we impose condition on solutions to be 
eigenfunctions of the number operator $\hat{N}$ and 
denote such functions by $\pc \equiv \lla x^a|n;c;j,m\rra$. 
We have the following system of equations
\begin{equation}
   \left\{\hat{N}, \hat{C}, \hat{J}^2, \hat{J}_{\!\ssc 12} \right\} \pc = \left\{n, \hbar^2(c^2-1) , \hbar^2 j(j+1 ), \hbar m \right\} \pc,
\end{equation}
where $\hat{C}= \frac{1}{2} \hat{J}_{\!\mu \nu} \hat{J}^{\!\mu \nu}$ 
is here represented by

\begin{equation} 
     \hat{C}= - \hbar^2 (x^a x_a) (\partial^a \partial_a) + 2 \hbar^2 x^a \partial_a
  + \hbar^2 (x^a \partial_a)^2
 \label{lam} \;.
\end{equation}

We make the following general coordinatization 
\begin{align}
     x^{\ssc 1}=&r \sqrt{1-u^2} \cf \st \;, &  x^{\ssc 2}=&r \sqrt{1-u^2} \sf \st \;,  \nonumber\\
    x^{\ssc 3}=&r\sqrt{1-u^2} \ct  \;,       & x^{\ssc 4}=& r u   \;, 
\label{coo}
\end{align}
where $x_a x^a= r^2$, hence $u$ has the range $[-1,1]$. 
A measure according to which the solutions have to be normalized is
\bea
d x = r^3 \sqrt{1-u^2} \st \,dr\,du \, d\theta\, d\phi\;.
\eea
We have $\pc \propto Y_{jm}(\theta,\phi)$, and assume 
factorization $\pc \propto R(r) U(u) Y_{jm}(\theta,\phi)$. 
We are left with two equations; one is 
$\hat{C} \pc = \hbar^2(c^2-1) \pc$,
and the other one is a number operator equation which 
can be casted simply as
\bea
- \hbar^2 \partial^a \partial_a \pc = \left( 2\hbar (\hat{N} +2)- x^a x_a\right) \pc
= \left( 2\hbar (n +2)-r^2 \right) \pc\;.
\eea

Expressed in terms of the new coordinates and combined, we get
\begin{gather}
\left(1-u^2\right) \frac{d^2U(u)}{du^2} -3u \frac{dU(u)}{du}
    + \left(c^2-1-\frac{j(j+1)}{(1-u^2)}\right) U(u)=0 \;,      
\label{Ueq}\\
\frac{d^2 R(r)}{dr^2}+ \frac{3}{r}\frac{dR(r)}{dr}  
    +\left(\frac{2\hbar(2+n)-r^2}{\hbar^2}  +\frac{1-c^2}{r^2}\right) R(r)=0 \;.                 
\label{Req}
\end{gather}
With the condition that the obtained functions are 
orthogonal, solutions are related to Legendre functions 
and Laguerre polynomials, 
\begin{align} 
U(u)&  = (1-u^2)^{-1/4} \left(a_{\ssc 1}\; P_{c-\frac{1}{2}}^{j+\frac{1}{2}}(u)+a_{\ssc 2}\;  
       Q_{c-\frac{1}{2}}^{j+\frac{1}{2}}(u) \right)\ ;, 
\label{Usol}\\
R(r)&\propto e^{-\frac{r^2}{2\hbar}}\left(\frac{r^2}{\hbar}\right)^{\!\frac{c-1}{2}}L_k^c \left(\frac{r^2}{\hbar}\right)\;,
 \end{align}
where\begin{equation} \label{k}
    k=\frac{n+1-c}{2} =0,1,2,...\;,
\end{equation} 
\emph{i.e.} normalizable solutions of (\ref{Req}) exist 
when $c=n+1,n-1,n-3 \dots$ with $k=0,1,2,\dots$
Further, to set the coefficients $a_{\ssc 1}$ and 
$a_{\ssc 2}$ in $U(u)$, given in  (\ref{Usol}), 
we compare solutions $\pc$ with the previously 
obtained Fock states wavefunctions (\ref{Herm}), upon 
applying the coordinate transformation (\ref{coo}), 
and deduce $a_{\ssc 1}=0$. Using the relation 
between Legendre functions (see \emph{e.g.}  \cite{htf}),
$Q_{c-\frac{1}{2}}^{j+\frac{1}{2}}\propto P_{c-\frac{1}{2}}^{-j-\frac{1}{2}}$,
we finally obtain normalized  solutions ($\int \psi^* \psi \,d^4x =1$)
\begin{equation}
    \pc =R_{nc}(r)U_{cj}(u)Y_{jm}(\theta, \phi)\;,    
\end{equation}
where
\begin{align}
Y_{jm}(\theta, \phi)
 =&\sqrt{\frac{(2j+1)}{4\pi}\frac{(j-m)!}{(m-m)!}} e^{i m \phi} P_j^m\left( \ct \right )\; ,
\nonumber\\
R_{nc}(r) =& e^{-\frac{r^2}{2\hbar}}\,\sqrt{\frac{2k!}{(k+c)!}}   \left(\frac{r^2}{\hbar}\right)^{\!\frac{c-1}{2}}
       L_k^c\left(\frac{r^2}{\hbar}\right),\qquad k=\frac{n+1-c}{2}\,
\nonumber\\
U_{cj}(u)=  &\sqrt{\frac{c(c+j)!}{(c-j-1)!}}    (1-u^2)^{-1/4}   P_{c-\frac{1}{2}}^{-j-\frac{1}{2}}(u) \;.
\end{align}
Particular linear combinations of such functions form 
the Fock state wavefunctions, \emph{i.e.}
 \begin{align}
 \lla r,u, \theta,\phi| n_{\ssc\! 1},n_{\ssc\! 2},n_{\ssc\! 3};n_{\ssc\! 4}\rra   
  = \!\!\!  \sum\limits_{c=n+1,n-1,...} \!\!\!\!\!\!  R_{nc}(r) \;   \sum\limits_{j=0}^{c-1}U_{cj}(u)
        \sum\limits_{m=-j}^j   Y_{jm}(\theta,\phi)    \; A_{cjm}^n \;. 
\label{psisol}
\end{align}
We can find the coefficients $A_{cjm}^n$ by comparison with 
$\lla x^a| n_{\ssc\! 1},n_{\ssc\! 2},n_{\ssc\! 3};n_{\ssc\! 4}\rra$, 
given in Eq.(\ref{Herm}). Explicit results for some of the lower
$n$ states are given in the appendix.

The above results match exactly to what we discuss
above in terms of the Cartesian tensors.  The 
$n$-level is a sum of irreducible representations of 
$c= n+1, n-1,...$, which are exactly the traceless 
irreducible tensors of rank $c-1$. Note that the $c$
value uniquely specifies the irreducible representation 
to which a basis state belongs. The $n$ value does not 
otherwise matter.  The $\left| n;1;0,0\right\rangle$ 
states for example, all transform exactly in the same 
way as the $\left| 0;1;0,0\right\rangle$ state, with
wavefunctions all of the form $f(x_ax^a)=f(r)$. 
Note that the Laguerre polynomial $L_k^c$ for our 
$k$ values is simply an order $k$ polynomial, hence 
$R_{nc}$ an order $n$ polynomial in $r$ times 
$e^{-\frac{r^2}{2\hbar}}$.
The pseudo-unitary inner product  among the 
$\left|n;c;j,m\rra$ states can be written as
\begin{equation}
    \lla\!\lla n';c';j',m'| n;c;j,m \rra\!\rra=(-1)^{n-j}\lla n';c';j',m'| n;c;j,m \rra
=(-1)^{n-j}\delta_{nn'}\delta_{cc'}\delta_{jj'}\delta_{mm'}\;.
\end{equation}
That is easy to appreciate as ${\mathcal P}_{\!\ssc 4}$
or simply  ${\mathcal P}{\mathcal P}_{\!\ssc (3)}^{-1}$ ,
where ${\mathcal P}$ is the full parity operator sending 
all $x^a$ to $-x^a$ and ${\mathcal P}_{\!\ssc (3)}$
the corresponding one for the  `3D' problem 
sending $x^i$ to $-x^i$ for which $\left|n;c;j,m\rra$
is an eigenstate with eigenvalues $(-1)^{n}$
and $(-1)^{j}$, respectively. 

\section{Discussions on Issues of Interpretations}\label{sec4}
Issues on the practical interpretation of the results 
are tricky. In lack of a solid practical setting that has 
been identified as to be depicted by a theory of the 
covariant harmonic oscillator, it is not quite possible 
to put the theory to be tested directly. Representing 
observables by non-Hermitian operators sure does 
not fit into the conventional interpretations of quantum 
theories, specifically in regard to the probability 
postulates. However, as stated above, it is not clear
at all that the usual probability notion should be
a part of a theory of wavefunctions over the 
`spacetime' variables. The formulation here has 
the position operators $\hat{X}_i$ and momentum
operators $\hat{P}_i$, and hence any observable 
corresponding to the function of those six basic 
observables, represented Hermitianly, in fact, in 
exactly the same standard way. Naively, that should 
include all physical observables, which would say that 
our formulation has no difficulty at all when applied 
to look at any of the physical observables at any 
specific value of the variable describing `time'. 
A bottom line, again, is that the pseudo-unitary
theory is fully unitary when the `time' variable for
the wavefunctions is restricted to a fixed value.
Hence, the usual probability interpretation in
connection to von Neumann measurements
performed at a definite time is not a problem.
The latter seems to be good enough for a theory 
of Lorentz covariant quantum mechanics 
interpreted along the usual perspective.
Taking up issues related to the nonhermitain time 
$\hat{X}_{\!\ssc 0}=ix_{\!\ssc 4}$ and energy 
$\hat{P}_{\!\ssc 0}= \hbar \partial_{x^{\ssc 4}}$ 
operators, some discussions about the notion of 
time in physical theories in Rovelli's book on quantum
gravity \cite{R} is very relevant, from which we would 
like to extract a few quotes here. For example, the 
author noted that time in Newtonian mechanics is
really an ``unobservable physical quantity", that 
it is enough for a theory to predict ``correlations 
between physical variables" but not necessarily 
values of the observables at any particular time.
Rovelli also observed that in general relativity,
 ``the coordinate time is not an observable", while
``dynamics cannot be expressed as evolution in 
$\tau$" (the proper time),  and that ``a fundamental 
concept of time may be absent in quantum gravity".
In view of all that, we can better consider the physics
picture of our time operator $\hat{X}_{\!\ssc 0}$. 
The first thing to note is that it should really be thought 
of more like the coordinate time. Taking quantum 
observables as noncommutative coordinates has 
been established as fully valid, for example in 
Ref.\cite{078} where it is shown how the six operators 
$\hat{X}_i$ and $\hat{P}_i$ can be seen as 
coordinates of a noncommutative symplectic 
geometry, which can alternatively be described as 
a commutative/real manifold of the projective 
Hilbert space, in the explicit language of a coordinate 
 transformation map. Our formulation of the
pseudo-unitary Lorentz covariant harmonic 
oscillator can be expected to fit well into the
Lorentz covariant generalization of that \cite{082}.
Note that without the notion of a noncommutative 
value for an observable \cite{079}, the quite intuitive
picture of the quantum phase space cannot be
made logically sound.  It is also relevant to note
that there has been a very substantial number of
studies on a plausible time operator in quantum
mechanics, though mostly not in a Lorentz covariant
theory, since the old days. A common conclusion 
from those studies is exactly that a Hermitian time
operator is not compatible with the theory.  On
the other hand, from the more mathematical
perspective, the observable algebra is essentially
agreed to be taken as a $C^*$-algebra, which 
corresponds to including all complex linear 
combinations of the physical/Hermitian operators. 
After all, a complex linear combination is in no 
sense any less `observable' than a real one. 
At first sight, having an energy operator $\hat{P}_{\!\ssc 0}$ 
to be nonhermitian posits a serious problem. However,
it is a mathematically unavoidable consequence of having
nonhermitian $\hat{X}_{\!\ssc 0}$. Upon a more careful
thinking, it is not at all clear that the  $\hat{P}_{\!\ssc 0}$ 
operator here has to be the physical energy as in the
usual quantum mechanics or classical special relativity.
In fact, it is easy to see that the $H_{\!\ssc R}(1,3)$
symmetry at the classical limit is still a symmetry bigger
than the Poincar\'e symmetry. The corresponding 
theory is certainly a theory more general than Einstein
special relativity, more like the so-called `parameterized
relativistic' theory (see for example Ref.\cite{F}). 
In fact, that kind of theory has essentially a translational
symmetry in the energy variable \cite{071}, which can
be seen as the usual notion of the physically indeterminate 
zero reference point of energy measurements,
in line with the notion of the energy or the energy 
operator as a coordinate variable.  

Another important point of view related to the idea 
of position and momentum operators as, actually
canonical, coordinate variables for the quantum
theory as symplectic dynamics is the symplectic 
geometric picture for the basic quantum mechanics 
(see for example Refs.\cite{078,CMP} and references 
therein) with the, infinite real dimensional, projective 
Hilbert space as the phase space. The Hamiltonian 
mechanics presents the dynamical theory well, at least 
when the measurement problem is not included. While 
the Copenhagen school framework with the probability 
picture gives a scheme to describe von Neumann 
measurements, it is hardly a dynamical/theoretical
description. The decoherence theory  \cite{de}, with
the statistical results from an open system perspective, 
we consider quite successful in that direction. In
principle, there is no fundamental difficulty in
formulating the latter equally successfully in the
symplectic geometric approach. The key point here is 
that a physical state, as a point in the corresponding
symplectic manifold, is completely unambiguous.  At 
least in principle, the state can be determined and 
the `full values'  of all observables, as known functions 
of the state \cite{078,079}, completely fixed 
accordingly. Such a  `full value' can be described 
as the noncommutative number \cite{079} which 
contains full information about the observable 
beyond the complete statistics of repeated von 
Neumann measurements. None of all that requires 
the observables to be Hermitian. In fact,  in the
noncommutative geometric picture the geometry 
is the dual object of the observable algebra, which
is basically the representation of the group 
$C^*$-algebra matching with  the quantum theory
as the representation of the basic/relativity symmetry
of $H_{\!\ssc R}(3)$ \cite{070}. We plan on going with 
the studies of a pseudo-unitary Lorentz covariant 
quantum theory along this line, with the latter generalized
to the  $H_{\!\ssc R}(1,3)$ symmetry, results and lessons 
from which would help us fully understand the physics 
of the covariant harmonic oscillator solutions given here.
It may be of interest to note further that from the
symplectic point of view, in terms of commutative or
noncommutative coordinates, dynamics is a specific
case of one-parameter Hamiltonian flows \cite{070}. 
The generator of the latter generally does not need to be 
inside the basic symmetry algebra. Even in the case of Galilean 
symmetry, the only case of the physical Hamiltonian being
included is the case of a free particle. Moreover, for 
the Lorentz covariant formulations, the evolution 
parameter should probably not be taken as the proper 
time. A parameter that corresponds to a proper time 
divided by the particle mass at the Einstein limit \cite{071},
as first introduced by Feynman back in 1950 \cite{Fe},
works better. With respect to the evolution parameter,
for the properly generalized  Lorentz covariant
Schr\"odinger equation, we are here only solving for 
what corresponds to the `time-independent',
 {\em i.e.} evolution parameter-independent 
covariant Schr\"odinger equation.

We can only sketch here above how the interpretational 
issues may be approached based on the known alternative 
perspectives. Beyond that, more studies within a full
setting of the Lorentz covariant quantum theory, still to
be explicitly formulated, have to be performed to help 
lighten up the physics picture. 

\section{Conclusions}\label{sec5}
A basic picture of the Fock states for the 
pseudo-unitary representation we presented 
here is more or less known. For the lack of 
interest from the conventional unitary quantum
theory line of thinking or otherwise, a detailed
analysis with comprehensive, consistent, explicit 
wavefunctions, the inner product, and the full
matching to the irreducible representations of
the $SO(1,3)$ has not been available. We present
here such a study.

The covariant harmonic oscillator problem in a
general setting of $SO(l,m)$ symmetry may serve 
as an important background for formulating the
corresponding quantum theory. It is all about an
irreducible representation of the $H_{\!\ssc R}(l,m)$
symmetry.  In fact, the authors came to the problem
with formulating such a covariant quantum theory
in mind. We see the un-conventional approach in the
direction of a pseudo-unitary representation  a
sensible one to explore, and the only reasonable
approach from a certain kind of background
perspectives.

Better appreciation of the physics
picture of  the theoretical framework could be
obtained with a full dynamical formulation of
such a quantum theory and, furthermore, by analyzing its 
application to various physical systems, especially the
experimentally accessible cases, like a motion of an 
electron under an electromagnetic field.

Looking carefully into the other theoretical 
applications of the Lorentz covariant harmonic
oscillator would of course also be useful though
the question of a solid practical setting for the
experimental applicability of such theories may
not be very well established. All that take more
efforts to which we hope to be able to contribute.
Our results here are given to provide the firm 
mathematical background for these kind of studies.

\newpage

\textbf{Appendix : List of explicit relations between some of
 the Fock states and the \boldmath $\left| n;c;j,m \rra$ basis states
of $SO(1,3)$ irreducible representations. }

\begin{align} \label{Acoeff}
 \left| n_{\ssc\! 1},n_{\ssc\! 2},n_{\ssc\! 3};n_{\ssc\! 4} \right\rangle=&\sum A_{cjm}^n\left| n;c;j,m \right\rangle  \nonumber \\
\left| 0,0,0;0 \right\rangle=&\left| 0;1;0,0 \right\rangle \nonumber\\
\left| 0,0,0;1 \right\rangle=&\left| 1;2;0,0 \right\rangle\nonumber\\
\left| 0,0,1;0 \right\rangle=&\left| 1;2;1,0 \right\rangle \nonumber\\
\left| 0,1,0;0 \right\rangle=&\frac{i}{\sqrt{2}}\left(\left| 1;2;1,1\right\rangle+\left| 1;2;1,-1\right\rangle\right) \nonumber\\
\left|1,0,0;0\right\rangle=&\frac{1}{\sqrt{2}}\left(\left| 1;2;1,-1\right\rangle-\left| 1;2;1,1\right\rangle\right)  \nonumber\\
\left| 0,0,0;2\right\rangle=&-\frac{1}{2}\left| 2;1;0,0\right\rangle+\frac{\sqrt{3}}{2}\left| 2;3;0,0\right\rangle \nonumber\\
\left| 0,0,2;0\right\rangle=&-\frac{1}{2}\left| 2;1;0,0\right\rangle-\frac{1}{2\sqrt{3}}\left| 2;3;0,0\right\rangle+\sqrt{\frac{2}{3}}\left| 2;3;2,0\right\rangle \nonumber\\
\left| 0,2,0;0\right\rangle=&-\frac{1}{2}\left| 2;1;0,0\right\rangle-\frac{1}{2\sqrt{3}}\left| 2;3;0,0\right\rangle-\frac{1}{\sqrt{6}}\left| 2;3;2,0\right\rangle-\frac{1}{2}\left( \left| 2;3;2,2\right\rangle+\left| 2;3;2,-2\right\rangle\right) \nonumber\\
\left| 2,0,0;0\right\rangle=&-\frac{1}{2}\left| 2;1;0,0\right\rangle-\frac{1}{2\sqrt{3}}\left| 2;3;0,0\right\rangle-\frac{1}{\sqrt{6}}\left| 2;3;2,0\right\rangle+\frac{1}{2}\left( \left| 2;3;2,2\right\rangle+\left| 2;3;2,-2\right\rangle\right) \nonumber\\
\left| 0,0,1;1 \right\rangle=&\left| 2;3;1,0 \right\rangle \nonumber\\
\left| 1,1,0;0 \right\rangle=&\frac{i}{\sqrt{2}}\left(\left| 2;3;2,-2\right\rangle-\left| 2;3;2,2\right\rangle\right) \nonumber\\
\left| 1,0,1;0 \right\rangle=&\frac{1}{\sqrt{2}}\left(\left| 2;3;2,-1\right\rangle-\left| 2;3;2,1\right\rangle\right) \nonumber\\
\left| 1,0,0;1 \right\rangle=&\frac{1}{\sqrt{2}}\left(\left| 2;3;1,-1\right\rangle-\left| 2;3;1,1\right\rangle\right) \nonumber\\
\left| 0,1,1;0 \right\rangle=&\frac{i}{\sqrt{2}}\left(\left| 2;3;2,-1\right\rangle+\left| 2;3;2,1\right\rangle\right) \nonumber\\
\left| 0,1,0;1 \right\rangle=&\frac{i}{\sqrt{2}}\left(\left| 2;3;1,-1\right\rangle+\left| 2;3;1,1\right\rangle\right) \nonumber\\
\left| 0,0,0;3\right\rangle=&\frac{1}{\sqrt{2}}\left(\left| 3;4;0,0\right\rangle-\left| 3;2;0,0\right\rangle\right)\;\nonumber
\end{align}

\noindent{\bf Acknowledgements \ }
We thank H.K. Ting for discussions.
O.K. is partially supported by research grant
number 107-2119-M-008-011
of the MOST of Taiwan.

\end{document}